# Storage and retrieval of a light in telecomband in a cold atomic ensemble

Dong-Sheng Ding,[†] Zhi-Yuan Zhou, Bao-Sen Shi,[*] Xu-Bo Zou, and Guang-Can Guo

*Key Laboratory of Quantum Information, University of Science and Technology of China, Hefei 230026, China*
Corresponding author: [†]dds@mail.ustc.edu.cn
[*]drshi@ustc.edu.cn

A telecom photon is a suitable information carrier in a fiber-based quantum network due to its lower transmission loss in fiber. Because of the paucity of suitable atomic system, usually the photon connecting different memories is in near infrared band, therefore the frequency conversion of the photon in and out of telecomband has to be required to realize the interface between the atomic-based memory and the photon-based carrier. In order for that, two atomic or other systems which could realize the frequency conversion have to be taken into account, and besides, one more atomic system as a storing media is need. So the ability of storing a photon in telecomband is an interesting and exciting topic. In this work, we give a first experimental proof of principle of storing a light in telecomband. The telecom light is directly stored and retrieved later through two nonlinear processes via an inverted-Y configuration in a cold atomic ensemble, therefore the interface between the memory and photon in other proposals is not needed. We believe our work may open a new avenue for long-distance quantum communication.

A possible quantum network consists of a kit in which quantum information could be stored and manipulated at will and a carrier by which different kits could be connected. Duan et. al. showed in their seminal paper [1] that an atomic system could be a suitable candidate for the memory and a photon could be a robust and efficient carrier due to its weak interaction with the environment. Owing to the paucity of suitable atomic system, the photon in near infrared band has to be acted as the carrier. It is well known that the photon in near infrared band is not suitable for long-distance transmission due to large loss in fiber. On the contrary, Ref. 2 clearly shows that the time needed to establish an entanglement between two remote legal users could be significantly reduced if a photon in telecom-band is used as an information carrier, compared with the case of a photon in near infrared band used. Therefore the frequency conversion of the photon in and out of telecomband by for example four-wave mixing [3-5] has to be required to realize the interface between the atomic-based memory and the photon-based carrier for long-distance communication. In order for that, two atomic or other systems which could realize the frequency conversion have to be taken into account. Besides, one more atomic system as a storing media is need. Therefore the whole system is complicated. So how to store a photon in telecomband directly is an exciting and interesting topic.

The strong demand for quantum memory has inspired new methodologies and led to experimental progresses for quantum storage using an atomic system via different mechanisms, for example, via electromagnetically induced transparency (EIT) [6, 7], atomic frequency combs [8, 9], Raman schemes [10, 11], and gradient echo memory [12, 13]. Another technique for storing a light is based on delayed four-wave mixing (FWM) process [3, 14] where the generated FWM field can be delayed and stored in atomic spin excitation. However, this method using a double-lambda configuration also works in the near infrared band, therefore it cannot be used to store the telecom light. In all above experiments, the photon carrying information is not in telecomband, i. e., there is no experimental demonstration of storing a light in telecomband, except the works reported by Gisin's group recently, where a light in telecomband is stored and retrieved later using photon echo technique in anerbium-doped solid system [15, 16]. In this work, we give a first experimental proof of principle that a light in telecomband could be stored and retrieved later by nonlinear processes in a cold atomic system. In our experiment, the atomic spin excitation contained the information of the light in telecomband is established between two atomic ground levels firstly through FWM in an inverted-Y configuration, then the atomic spin excitation is converted back to the light in telecomband later through a two-photon absorption process. The efficiency of whole process is about $1.5 \times 10^{-7}$ at present. Compared with other works, the interface between the memory and photon in other proposals is not needed. Our result may open a new avenue to long-distance quantum communication.

Two experiments are reported in this work: In the first, we convert a light in telecomband into an atomic spin excitation through FWM, and then convert the atomic spin excitation to a near infrared light through EIT in a Lambda configuration. In the second, the atomic spin excitation built up through FWM is converted back a light in telecomband through a two-photon absorption process. A cigar-shaped atomic cloud of $^{85}$Rb atoms, trapped in a two-dimensional magneto-optical trap (MOT), was used as the storing media. The size of cloud is about $30 \times 2 \times 2 mm^3$. The total atom number is $9.1 \times 10^8$ [17]. The inverted-Y type configuration used in our experiment is shown in Fig. 1(a). It consists of two degenerated ground states |1> and |2> ($5S_{1/2}$ F=3); one intermediate state |3> ($5P_{1/2}$ F'=2) and one upper state |4> ($4D_{3/2}$ F''=3). The transition frequency between the ground state and the intermediate state matches to the D2 line (795 nm) of atom $^{85}$Rb, and the transition between the intermediate state and the upper state can be driven by an infrared laser at 1475.6 nm. We experimentally generate a new field at 795 nm with the combinations of a laser beam at 795 nm and two laser beams at 1475 nm via a non-collinear FWM process. In this process, all lasers are resonant with correspond atomic transitions because in this case the nonlinear gain is large [18]. Based on this process, we carried on the experiment of storing and retrieving a light in telecomband. The simplified experimental setup is shown in Fig. 1(c). A cw (continuous wave) laser beam at 795 nm from an external-cavity diode laser (DL100, Toptica) is input to the cold atomic cloud as the coupling field. A cw laser beam at 1475.6 nm from another external-cavity diode laser (DL100, Prodesign, Toptica) is divided into two fields by a beam splitter: the pump and signal fields. These fields are horizontal polarization and modulated by a +80 MHz acousto-optic modulators (AOM) to generate an optical pulse sequence. The up-converted field at 795 nm with the $\sigma^+$ polarization is monitored using a high-speed camera (1024×1024, iStar 334T series, Andor) or a PMT (Hamamatsu, H10721). The coupling field with the $\sigma^-$ polarization acts as a pumping field in FWM process and as a reading light in the retrieval process. Our experimental image-detecting is based on a 4-f imaging system, which consists of two lenses with 500 mm focus length.

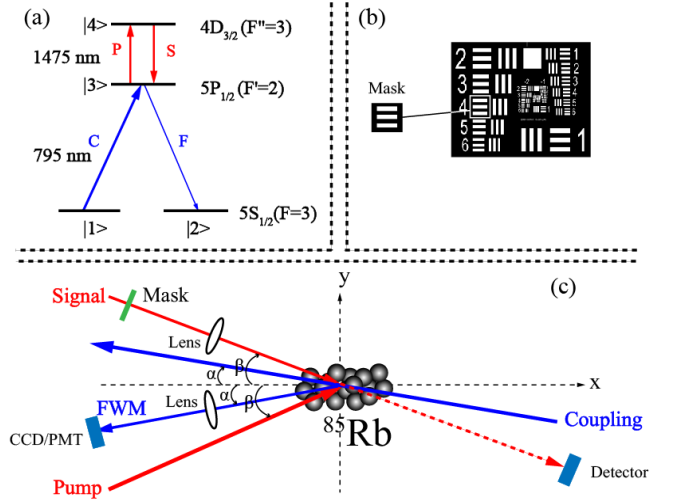

Fig. 1 (a) The experimental energy diagram. Two lasers with different wavelengths of 795 nm and 1475 nm are used to perform the experiment. The two degenerated ground states are the sublevels of ($5S_{1/2}$ F=3) $m_F$=-1, $m_F$=-3. The experimental mask is shown by Fig. 1(b), which is the standard imaging mask. (c) The experimental setup. The angles between the pump, coupling and signal fields are α≈1°, β≈2° respectively.

Before showing the experimental results, we give a simple theoretical analysis firstly. The effective Hamiltonians for our system is below:

$$H_{eff} = -\hbar(\Omega_c |3\rangle\langle 1| + \Omega_F |3\rangle\langle 2| + \Omega_p |4\rangle\langle 3| + \Omega_s |4\rangle\langle 3|) + H.C . \quad (1)$$

Where, $\Omega_p$, $\Omega_s$, $\Omega_c$ and $\Omega_F$ are the Rabi frequencies of the pump, signal, coupling and the generated FWM fields. The density matrix equation of motion is

$$\dot{\rho}_{ij} = -\frac{i}{\hbar}\sum_k (H_{ik}\rho_{kj} - \rho_{ik}H_{kj}) - \Gamma_{ij}\rho_{ij}. \quad (2)$$

where, $\Gamma_{ij}$ ($i \neq j$) describes the complex decay rate from |i> to |j>, $\Gamma_{ij}$ ($i=j$) describes the decay rate of $\rho_{ii}$. In our system, the state |3> is degenerate, the decay rate is 2γ. The decay rate of state |4> is 2Γ. We consider the zero-order perturbation expansion with the assumption of $\Omega_p$, $\Omega_s$, $\Omega_c$ >> $\Omega_F$, and derive the steady-state solutions of density matrix equation on the condition of $\rho_{44}=\rho_{33}=0$. The atomic spin

excitation $\rho_{12}$ expression obtained is:

$$\rho_{12} = \frac{\gamma\Omega_F[i(\Omega_p+\Omega_s)^2+\gamma\Omega_c^*]}{-\gamma^2\Omega_c^2+(\Omega_p+\Omega_s)^4} \quad (3)$$

If the pump and signal fields don't exist, the Eq. (3) can be simplified to be the general expression of $\rho_{12}=-\Omega_F/\Omega_c$, which approximately describes the storage of light in a lambda type configuration. When we turn off the coupling, pump and signal fields, the density of operator $\rho_{12}$ follows the time $t$ with exponential decay.

$$\rho_{12}(t) = \frac{\gamma\Omega_F[i(\Omega_p+\Omega_s)^2+\gamma\Omega_c^*]}{-\gamma^2\Omega_c^2+(\Omega_p+\Omega_s)^4} e^{-\frac{t}{\tau}} \quad (4)$$

where $\tau$ is the coherence time determined by the decay between two ground states. The atomic spin excitation state $\rho_{12}(t)$ contains the spatial information if the signal field is imprinted spatial information. When the pump and the coupling fields or only coupling field are turned on later, the signal field is retrieved through the previous established atomic spin excitation.

We firstly performed the nonlinear FWM process to establish the atomic spin excitation between the two ground states. The powers of signal, pump and coupling fields are 0.20 mW, 0.63 mW and 1.16 mW respectively. The generated FWM signal at 795 nm monitored by a PMT is shown in Fig. 2. When we turned off the coupling and pump fields simultaneously, the generated FWM signal at 795 nm is converted to the atomic spin excitation between two ground states and stored. After a while, when we switch the coupling field on, the generated FWM signal is retrieved. The Fig. 2 shows the experimental result, where the black dots represent the generated FWM field and the red dots are the stored FWM field. The storage time is about 3.4 μs. In this process, we clearly demonstrate that the generated FWM field can be stored and retrieved by using an inverted-Y type configuration.

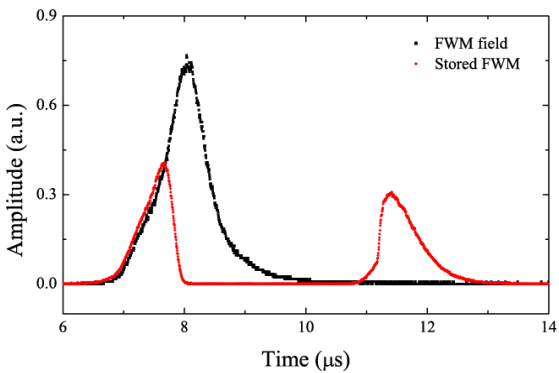

Fig. 2 The black dots represent the generated FWM signal. The red dots are the leaked and retrieved FWM signals. The storage time is about 3.4 μs.

In order to clearly show whether the information contained in light in telecomband is converted to the atomic spin excitation, the signal light is imprinted a spatial information with a standard mask (USAF target) shown by Fig. 1(b), and we want to see whether the spatial information contained by telecomband light could be stored and retrieved. We used the high efficiency camera triggered by a delay generator (DG535) to monitor the generated FWM signal. By using the 4-f imaging system, we can obtain the images of leakage and retrieval, which are shown in Fig. 3, where(a) is the leaked image, and (b) is the retrieval image. By plotting the intensity profiles in the vertical direction through the centers of image, we calculated the visibility of the retrieved images using the formula V= $(I_{max}-I_{min})/(I_{max}+I_{min})$, where $I_{max}$ and $I_{min}$ are the maximal and minimal intensities along the vertical direction. The calculated visibilities of leaked image and retrieved image are 0.83 and 0.67 respectively. The experimental fact of the spatial information imprinted in telecomband light being stored and retrieved later clearly verified that the information imprinted in telecom light is converted to the atomic spin excitation like descriptions of Eq. (4).

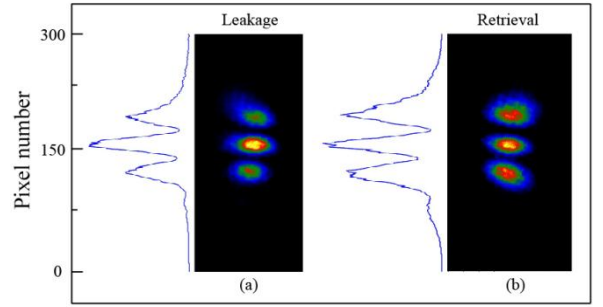

Fig. 3 (a) The leaked image of FWM signal. (b) The retrieved image of FWM signal.

Now we move to the main topic of this work: storing a telecom light and retrieving it later. In order to obtain the retrieved signal, we must turn on the coupling and pump fields simultaneously again. We make the periods of the signal, pump and coupling pulses to be 10 μs, the widths of each signal and coupling pulse are set about 1 μs and 7 μs respectively. In our system, the back-edge of each pump and coupling pulse writes the signal into the atomic collective spin excitation state, whereas the front-edge of the following pulse is used to read out the stored signal pulse from the atoms. This design can significantly improve the strength of the retrieved signals. We carried out our experiment at a repetition frequency of 100 Hz. Each iteration consisted of a loading period of 9.7 ms and an experimental window of 300 μs that accommodated 20 signal, pump and coupling pulses. Therefore, the total number of probe pulses per second was 2000, resulting in a high signal-to-noise ratio (SNR) retrieved signal. The timing sequence of the measurement is shown in Fig. 4.

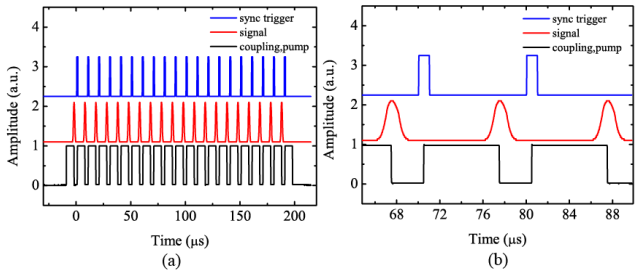

Fig. 4. The time sequence of measurement. The blue line is the syns. trigger signal, the red line is the signal pulses. The black line represents the coupling and pump optical pulses. (a) is the time duration of one experiment at a repetition frequency of 100 Hz. (b) is the enlarged picture in time domain.

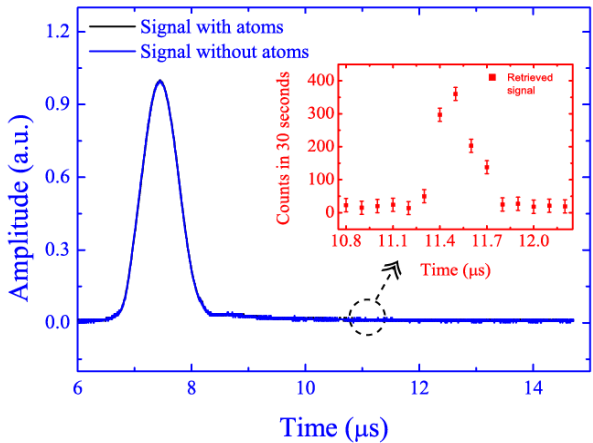

Fig. 5 The leaked signal and the retrieved signal in time domain. The error bars are statistics results. The storage time is about 4 μs. The leaked signal is measured by a home-made detector because of its strong intensity. The retrieved signal is detected by a single photon detector because of its weak intensity.

We used a home-made infrared detector to detect the intensity of the leaked signal field in time domain because of its strong intensity, the leaked signal is obtained shown by blue part of Fig. 5. An infrared detector (ID Quantique, InGaAs Photon Detector with 8% detection efficiency) is used to detect the retrieved signal in telecomband because of its weak intensity. This detector works in gated mode and is triggered by a sync signal generated by an arbitrary function generator (AFG3252). The delay time of sync signal can be adjusted by this arbitrary function generator itself. When we make the sync signal cover the retrieved signal completely, the intensity information could be obtained by triggering the detector. Through changing the delay time step by step, the time resolution intensity information of retrieved signal field could be obtained, which is shown by the red part of Fig. 5. In the data processing, we subtract the noises. The noises are mainly from 1: the scattered photons of the pumping laser, which are near equal to true signal counts; 2: the spontaneously Raman emitted photons along the signal direction. We find they are very small and can be safely omitted compared to the intensity of retrieved signal. In our experiment, the photon number per pulse of signal field is about $10^8$, the efficiency of whole process is estimated to be on the order of $10^{-7}$. The low efficiency in this memory process is mainly due to the small nonlinear gain in our system. From our experimental results, we estimated the efficiency of FWM process in our system to be on the order of $10^{-3}$. The efficiency of storage may be increased by for example improving atomic optical density, optimizing some experimental parameters such as Rabi frequency and detuning of all fields etc. Besides, in our system, the no-collinear configuration is used to reduce the noise caused by the scattering of strong lasers, but it somehow reduces the efficiency.

In the experiment of storing a light in telecomband, the mask is removed, therefore no spatial information is included. This is mainly due to the lack of a suitable infrared-band CCD camera worked at single-photon level in our Lab. We believe that a light imprinted an image in telecomband could be stored in our system definitely. In the experiments reported here, we keep storage time relatively short in order to curtail the overall duration of the experiments, so that a relatively high SNR and a clear image could be achieved.

One big advantage of this scheme is the realization of the interface between the memory and the carrier and the storage of information carried by the light in only an atomic ensemble, which usually are realized with three atomic ensembles in other proposals, because one ensemble is needed to realize the frequency upconversion of the photon in telecomband to match the atomic transition wavelength, the second is used to store the upconverted photon, and the third to down-convert back the light in telecomband for long-distance transmission. Another thing we want to mention is that although our experiment was done in a cold atomic system, we believe that the scheme could be performed in a hot atomic ensemble too, therefore the whole system could be simplified significantly.

In summary. The first experimental demonstration of storing a light in telecomband is reported. In this experiment, the telecom light is directly stored in a cold atomic ensemble and retrieved later through two nonlinear processes. Although the efficiency of our proof of principle is relatively low, it may open a new avenue for long-distance quantum communication.

**Acknowledgements**

We thank Dr. Wei Chen and Dr. Bi-Heng Liu for kindly loan us single photon detector and for other technique support. This work was supported by the National Natural Science Foundation of China (Grant Nos., 11174271, 61275115), the National Fundamental Research Program of China (Grant No. 2011CB00200), and the Innovation fund from CAS, Program for NCET.

**Author contributions**

BSS and DSD conceived the experiment for discussion. The experimental work and data analysis were carried out by DSD and BSS, with assistance from ZYZ. BSS and DSD wrote this paper. BSS, XBZ and GCG supervised the project.

**References**

1. Duan, L.-M., Lukin, M. D., Cirac, J. I. and Zoller, P., Long-distance quantum communication with atomic ensembles and linear optics, Nature **414**, 413–418 (2001).
2. R. T. Willis, "Photon Pair Production from a Hot Atomic Ensemble in the Diamond Configuration," Ph. D. thesis, University of Maryland, College Park, (2009).
3. Ryan M. Camacho. Praveen K. Vudyasetu. John C. Howell. Four-wave-mixing stopped light in hot atomic rubidium vapour, Nature Photonics 3, 103-106 (2009).
4. Ding, Dong-Sheng, Zhou, Zhi-Yuan, Shi, Bao-Sen, Zou, Xu-Bo and Guo, Guang-Can, Image transfer through two sequential four-wave-mixing processes in hot atomic vapor, Phys. Rev. A. 85, 053812 (2012).
5. Ding, Dong-Sheng, Zhou, Zhi-Yuan, Shi, Bao-Sen, Zou, Xu-Bo and Guo, Guang-Can, Experimental up-conversion of images, Phys. Rev. A. 86, 033803 (2012)
6. Wu, Jinghui, Ding, Dongsheng, Liu, Yang, Zhou, Zhiyuan, Shi, Baosen, Zou, Xubo and Guo, Guangcan, Storage and Retrieval of an Image using Four-Wave Mixing in a Cold Atomic Ensemble, arxiv:1204:0955v4.
7. Fleischhauer, M. and Lukin,M. D., Dark-State Polaritons in Electromagnetically Induced Transparency, Phys. Rev. Lett. 84, 5094 (2000).
8. Phillips, D. F., Fleischhauer, M., Mair, A., Walsworth, R. L., and Lukin, M. M., Storage of Light in Atomic Vapor, Phys. Rev. Lett. 86, 783 (2001).
9. Afzelius, M., Simon, C., de Riedmatten, H. and Gisin, N., Multimode quantum memory based on atomic frequency combs, Phys. Rev. A., 79, 052329 (2009).
10. de Riedmatten, H., Afzelius, M., Staudt, M. U., Simon, C., and Gisin.N.,A solid-state light–matter interface at the single-photon level, Nature, 456, 773 (2008).
11. Rein, R. F., Michelberger, P., Lee, K. C., Nunn, J., Langford, N. K., and Walmsley, I. A., Single-Photon-Level Quantum Memory at Room Temperature, Phys. Rev. Lett. 107, 053603 (2011).
12. K. F. Reim, J. Nunn, V. O. Lorenz, B. J. Sussman, K. C. Lee, N. K. Langford, D. Jaksch and I. A. Walmsley, Towards high-speed optical quantum memories, Nat. Photon. 4, 218 (2010).
13. Hosseini, M., Sparkes, B. M., Campbell, G., Lam, P. K. and Buchler, B. C., High efficiency coherent optical memory with warm rubidium vapour, Nature Comm. 2, DOI: 174 (2011).
14. Hosseini, M., Campbell, G., Sparkes, B. M., Lam, P. K. and Buchler, B. C., Unconditional room-temperature quantum memory , Nature Phys. 7, 794 (2011).
15. M. U. Staudt, M. Afzelius, H. de Riedmatten, S. R. Hastings-Simon, C. Simon, R. Ricken, H. Suche, W. Sohler, and N. Gisin, Phys. Rev. Lett. 99, 173602 (2007).
16. BjörnLauritzen, Jirˇı´ Mina´rˇ, Hugues de Riedmatten, Mikael Afzelius, Nicolas Sangouard, Christoph Simon, and Nicolas Gisin, PRL 104, 080502 (2010)
17. Liu, Y., Wu, J.-H., Shi, B.-S., and Guo, G.-C., Realization of Two-Dimensional Magneto-Optical Trap with a High Optical Depth, Chin. Phys. Lett. 29, 024205 (2012).
18. M. D. Lukin, P. R. Hemmer, M. O. Scully, Resonant nonlinear optics in phase-coherent mediaAdv. At. Mol. Opt. Phys. 42, 347 (2000).